# Proactive Software Supply Chain Risk Management Framework (P-SSCRM) Version 1.01 April 28, 2025

**Laurie Williams**
**North Carolina State University**
**laurie_williams@ncsu.edu**

**Sammy Migues**
**Imbricate Security**
**sammy.migues@gmail.com**

**Jamie Boote**
**Ben Hutchison**
**Black Duck**
**{jboote, hutchiso}@blackduck.com**

Acknowledgments

The development P-SSCRM was funded by Synopsys. The authors thank all of the organizations and individuals who participated in assessment interviews that led to the refinement of the P-SSCRM framework. The following individuals provided input and guidance through the framework creation process: Chris Madden, Rob Hines, and DJ Schleen of Yahoo; and Karen Scarfone of NIST.

**Provide input**

Submit comments on this publication to Laurie Williams, laurie_williams@ncsu.edu.

Table of Changes

| Version | Summary of Changes |
| --- | --- |
| 1.01 | • Typos fixes<br>• Change affiliation of authors Jamie Boote and Ben Hutchison<br>• Change to Creative Commons License CC BY-SA 4.0 Attribution-ShareAlike 4.0 International |

# 1. P-SSCRM Introduction and Background

Software organizations largely did not anticipate how the software supply chain would become a deliberate attack vector. The software industry has moved from passive adversaries finding and exploiting vulnerabilities in code contributed by well-intentioned developers, such as log4j[1], to a new generation of software supply chain attacks, where attackers also aggressively implant vulnerabilities directly into dependencies (e.g., the protestware of node-ipc[2]). Adversaries also find their way into builds and deployments, such as with SolarWinds[3], to deploy rogue software. Once implanted, these vulnerabilities become an efficient attack vector for adversaries to gain leverage at scale by exploiting the software supply chain. The rapid growth in software supply chain attacks has driven governments and organizations to take deliberate action to reduce software supply chain risk.

The Proactive-Software Supply Chain Risk Management (P-SSCRM) Framework described in this document is designed to help you understand and plan a secure software supply chain risk management initiative. P-SSCRM was created through a process of understanding and analyzing real-world data from nine industry-leading software supply chain risk management initiatives as well as through the analysis and unification of ten government and industry documents, frameworks, and standards. Although individual methodologies and standards differ, many initiatives and standards share common ground. P-SSCRM describes this common ground and presents a model for understanding, quantifying, and developing a secure software supply chain risk management program and determining where your organization's existing efforts stand when contrasted with other real-world software supply chain risk management initiatives.

## Where did the P-SSCRM come from?

The Proactive Software Supply Chain Risk Management (P-SSCRM) Framework results from a unique study of real-world software supply chain risk management initiatives and the union of the tasks in ten government and industry documents (standards and frameworks). Tasks in the P-SSCRM are mapped to one or more of these standards and frameworks. We present the model as built directly from these tasks and from data observed in real-world software supply chain risk management initiatives from a diverse and global collection of firms through data collected in 2022 and 2023.

The ten frameworks used in the foundation and mapping of P-SSCRM tasks and their mapping references (in parentheses) are:

1. Executive Order 14028 (EO)
2. NIST Secure Software Development Framework version 1.1 (800-218) (SSDF)
3. NIST Cybersecurity Supply Chain Risk Management Practices for Systems and Organizations (800-161r1), only the subset of tasks specifically identified in this document as mapping back to the Executive Order (EO) (800-161)
4. DHS/CISA Secure Software Self-Attestation Common Form (Self attestation)

[1] https://nvd.nist.gov/vuln/detail/CVE-2021-44228
[2] https://nvd.nist.gov/vuln/detail/cve-2022-23812
[3] https://nvd.nist.gov/vuln/detail/CVE-2020-10148

5. Building Security In Maturity Model Version 13 (BSIMM)
6. Supply-chain Levels for Software Artifacts v1.0 (SLSA)
7. OpenSSF Secure Supply Chain Consumption Framework (S2C2F)
8. Open Web Application Security Project Software Component Verification Standard Version 1.0 (OWASP SCVS)
9. Cloud Native Computing Foundation – Software Supply Chain Best Practices (CNC SSC)
10. OpenSSF Scorecard metrics (OSSF Scorecard)

## What is the P-SSCRM's purpose?

The P-SSCRM is a holistic framework that an organization can use to proactively mitigate software supply chain risk through guided adoption of tasks; and that supports assessment, scoring, and comparison against industry peers, standards, and guidelines. The P-SSCRM contextualizes and quantifies the tasks contained across multiple standards and frameworks to those carried out by various kinds of organizations.

As various standards and organizational initiatives use different methodologies and terminology, the P-SSCRM provides a framework that enables a uniform description of software supply chain risk management initiatives. Our P-SSCRM framework and task descriptions provide a common vocabulary for explaining the salient elements of a software supply chain risk management initiative, thereby allowing a comparison of initiatives that use different terms, operate at different scales, exist in different parts of the organizational chart, operate in different vertical markets, or create different work products.

We created the P-SSCRM to learn how software supply chain risk management initiatives work and to provide a resource for people looking to create or improve their own software supply chain risk management initiative. In general, every firm creates its software supply chain risk management initiative with some high-level goals in mind. The use of the P-SSCRM framework is appropriate if your business goals for software security include:

- Informed risk management decisions
- Clarity on what is “the right thing to do” for a holistic set of roles involved in software supply chain security based upon the guidance referenced in the framework
- Improved software and associated supply chain security and compliance assurance.

The P-SSCRM framework provides the structure for a “descriptive” model. That is, P-SSCRM is not a prescriptive model that recommends what an organization should do to reduce software supply chain risk. Instead, P-SSCRM provides information on what the organizations that have undergone a P-SSCRM assessment are doing. Put another way, P-SSCRM is not a set of best practices defined by some committee for some one-size-fits-all generic problem. Rather, P-SSCRM is a set of actual practices being performed daily by forward-thinking firms.

## Terminology & Definitions

Below, we provide definitions for the P-SSCRM terms introduced and used in this framework:

- **Secure SDLC (S-SDLC).** A software lifecycle with integrated software security checkpoints and activities.
- **Software Supply Chain Risk Management Initiative (SSCRM-I).** An organization-wide program to reduce software supply chain risk activities in a coordinated fashion.
- **Task.** Actions or efforts conducted in the process of implementing a secure software application and of reducing the security risk of that application and for its producing organization. Each task has a lower-level objective to aid in secure software development and risk reduction. For example:
  - ***P.3.1 Component and container choice:*** *make informed third-party component and container choices*
- **Practice.** A grouping of P-SSCRM tasks that have a similar mid-level objective to aid in secure software development and risk reduction. The 73 tasks of the P-SSCRM are organized into 15 practices. For example,
  - **P.3 Manage component and container choices:** software supply chain risk can be reduced by careful choice and handling of third-party components and containers.
- **Group.** A grouping of P-SSCRM practices with a similar high-level objective to aid in secure software development and risk reduction. The 15 practices of the P-SSCRM are organized into four groups: Governance, Product, Environment, and Deployment.
  - **Product (P):** Tasks to lead to deploying a secure product with minimal vulnerabilities with associated required attestations and artifacts.

Below, we define the P-SSCRM roles who conduct the tasks to reduce software supply chain security risk:

- **Business Manager:** This grouping of roles includes compliance, risk, and vendor managers.
- **Architect/Developer:** This role designs, implements, and tests a software product.
- **Information Technology (IT):** This role provides the hardware, software, and services infrastructure to enable an organization to receive, store, retrieve, transmit, and manipulate data.
- **DevOps:** This role provides the technology for deploying, delivering, and operating software applications and services through the integration and collaboration between development teams and operations teams.
- **Software Security:** This role creates and facilitates processes and procedures for secure software development at an organizational level.

## 2. The Proactive Software Supply Chain Risk Management (P-SSCRM) Framework

Figure 1 shows the structure of the Proactive Supply Chain Risk Management (P-SSCRM) Framework. It includes four broad groups of Governance, Product, Environment, and Deployment. Our P-SSCRM as well as both practice and task descriptions, provide a common vocabulary for explaining the salient elements of an SSCRM-I. Within the four P-SSCRM groups are 15 practices (e.g., Perform compliance). The current version of the P-SSCRM, the P-SSCRM1, is composed of 73 software supply chain risk management tasks that are organized into these 15 practices.

| P-SSCRM Model (4 Groups, 15 Practices, 73 Tasks) | | | |
|---|---|---|---|
| Groups | | | |
| Governance (23 tasks) | Product (19 tasks) | Environment (23 tasks) | Deployment (8 tasks) |
| The Governance Group contains 5 Practices, made up of Tasks that focus on the organization and measurement of a secure software supply chain and of policies for decision making, accountability to third-party obligations, and remaining compliant with legal and regulatory requirements. | The Product Group contains 5 Practices, made up of Tasks to lead to the deployment of a secure product with minimal vulnerabilities with associated required attestations and artifacts. | The Environment Group contains 3 Practices, made up of Tasks to protect the confidentiality and integrity of source code, software components, and the build infrastructure from tampering and unauthorized access. | The Deployment Group contains 2 Practices, made up of Tasks for identifying, analyzing, and addressing vulnerabilities in products. |

**Figure 1: P-SSCRM Framework of Four Groups**

Figure 2 displays the P-SSCRM groups and practices in the context of a product lifecycle model, annotating the primary role responsible for each practice. The practices and associated tasks protect the integrity of source code, the build environment, deployed and running software applications. They also include practices to securely decommission a software product at its end of life. The practices that appear in solid boxes along the top and left side of the lifecycle indicate practices that are done throughout the product lifecycle.

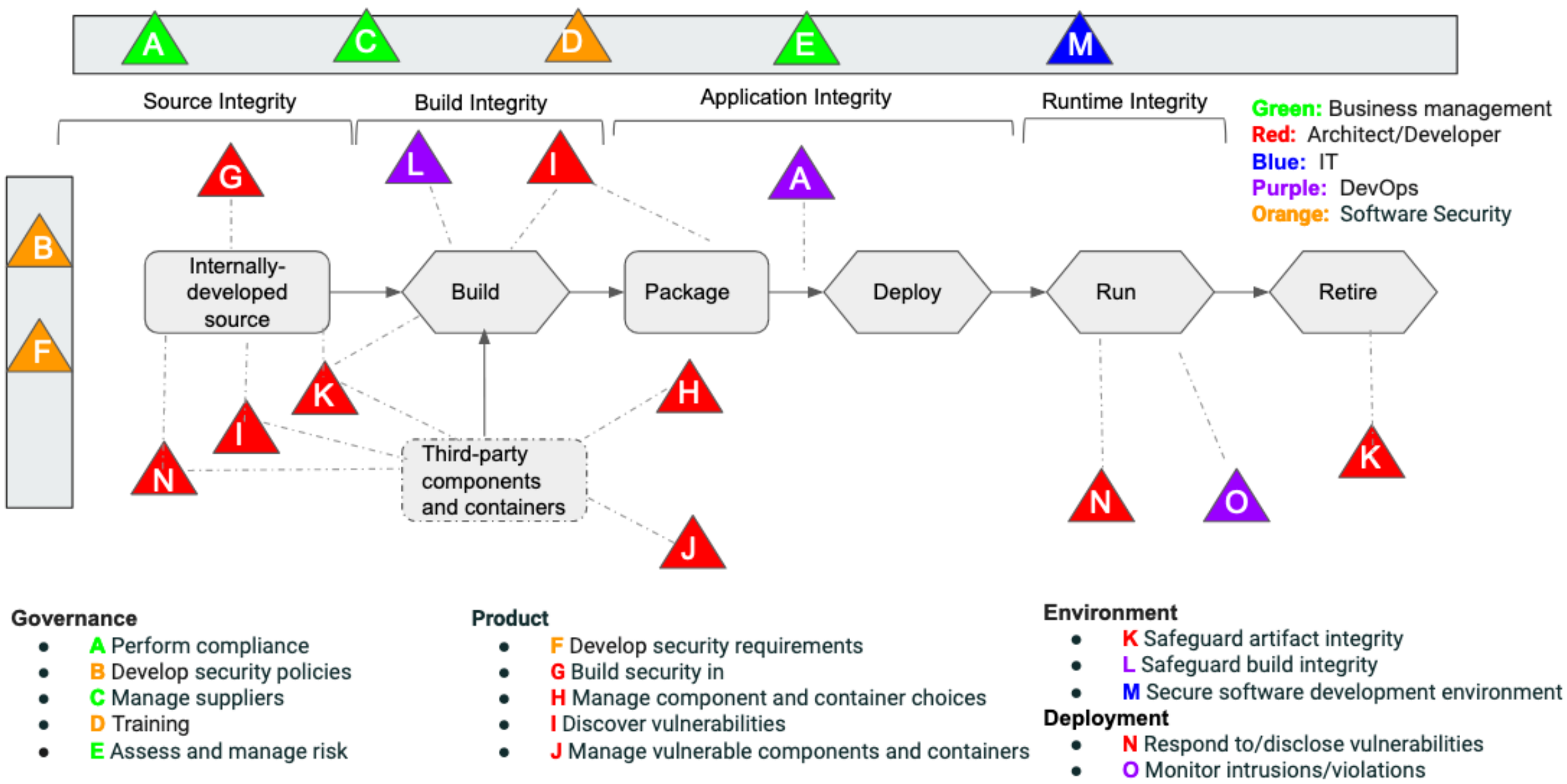

**Figure 2: P-SSCRM Lifecycle Model by Role**

Figure 3 displays an example of one of the 73 P-SSCRM tasks. The example task is in the Governance group (G) and in the Perform Compliance (G.1) practice. The task, Organizational Security Requirements (G.1.1), has a unique identifier, a description of the actions a business manager should take, assessment questions, and a mapping to the six frameworks that prescribe this task. The green triangle A for the Perform Compliance practice in Figure 2 indicates that a business manager does task G.1.1 throughout the lifecycle.

| | | | | |
|---|---|---|---|---|
| **GOVERNANCE (G):** Tasks that focus on the organization and measurement of a secure software supply chain and of policies for decision making, accountability to third-party obligations, and remaining compliant with legal and regulatory requirements. | | | | |
| **G.1 Perform compliance:** Compliance is following established guidelines or specifications or becoming so, possibly through a demonstration or audit. | | | | |
| G.1.1 Org security requirements | Organizational security requirements, such as those imposed by standards and regulations, are included in the SDLC. | Identify, document, communicate, and maintain security requirements and policies for the organization's software development infrastructure and secure SDLC. Maintain the requirements and policies over time. Incorporate constraints imposed by standards and regulations and customer-driven security requirements. | Do you have a defined secure SDLC that the engineers are aware of? Do you define security requirements and policies for the organization, its development infrastructure, contributions, and processes? How are these requirements and contributions maintained over time? Are constraints imposed by regulatory and compliance drivers included in these requirements, policies, and the SDLC? | **EO:** 4e(ix)<br>**SSDF:** PO.1.1<br>**BSIMM:** CP1.1, CP1.2, CP1.3, SR1.1, SR2.2, SR3.3<br>**800-161:** SA-15<br>**CNCF SSC:** C: Establish and adhere to contribution policies<br>**Self-attestation:** 2 |

**Figure 3: Example P-SSCRM Task**

## How should I use the P-SSCRM?

The P-SSCRM can be used as a measuring stick for secure software supply chain risk management program initiatives. You can then identify goals and objectives of your own and look to the P-SSCRM to determine which further tasks make sense for you and your organization.

While there's been a steady trickle of software supply chain attacks since at least 2017, the real growth in attacks and the resulting focus on software supply chain risk management, has only occurred over the last three years. Instilling software supply chain risk management into an organization takes careful planning and always involves broad organizational change. Using the P-SSCRM as a guide for your software supply chain risk management initiative, you can leverage the many years of experience captured in the frameworks authored by reputable organizations and the experience of the nine industry-leading software supply chain risk management initiatives. You should tailor the implementation of the activities of the P-SSCRM to your organization (carefully considering your objectives). Note that no organization surveyed during this work carries out all the tasks described in the P-SSCRM.

## WHO SHOULD USE THE P-SSCRM?

The P-SSCRM is appropriate for anyone responsible for creating and executing a secure software supply chain risk management initiative or looking to incorporate a higher degree of software security assurance throughout their existing program. Five example P-SSCRM user stories are also presented below:

- As an **individual contributor (business manager, software developer or tester, information technology, DevOps engineer**), I want to understand the tasks I should adopt to reduce software supply chain risk.
- As a **CISO or senior security engineer**, I want to guide the adoption of secure software supply chain tasks in well-defined iterations to reduce security risk.
- As a **CTO/CIO**, I want to demonstrate concrete improvements to software supply chain security.
- As a **C-level executive**, I want to compare my organization's secure software supply chain tasks with industry trends and to evaluate my organization's tasks with those prescribed in industry standards, guidelines, and publications.
- As an **auditor or legal counsel**, I want to assess secure software supply chain practices throughout an organization and for (self-) attestation to reflect conformance accurately.

## WHY WERE TEN FRAMEWORKS INCLUDED?

The ten frameworks were chosen because government and industry practitioners frequently talked about all ten at meetings and summits and on Slack and blogs during the months of development of P-SSCRM. As the tasks of each framework were added to P-SSCRM, each task was analyzed for bi-directional equivalence with an existing P-SSCRM task. Two tasks are bi-directionally equivalent if they have the same meaning but most likely different wording/phrasing in their definition in the different frameworks. A mapping was added when a task was considered equivalent to an existing task, and a new task was created otherwise.

Some of the mappings between tasks are based on mappings contained in an original framework:

- Tasks in the NIST 800-218 (SSDF) were mapped to the EO, BSIMM, OWASP SCVS, and NIST 800-161.
- Tasks in S2C2F were mapped to NIST 800-161, NIST 800-218 (SSDF), OWASP SCVS, and CNCF SSC.
- Tasks in self-attestation were mapped to NIST 800-218 (SSDF).

We did the other mappings by reading the descriptions of the tasks in the various documents to assess bi-directional equivalence. The mappings were freely distributed to interested parties for feedback over a six-month period.

As each of the ten standards was considered for inclusion in the P-SSCRM, the strengths of each and their value in working together synergistically to provide a holistic view of software supply chain risk reduction by all five roles were realized. The number of tasks in each of the four groups that came from each framework is shown in Table 1.

The bolded numbers in Table 1 indicate the framework most influential in providing the tasks for a group. Twenty of the 23 Governance tasks came from the NIST 800-161 framework. The Business Manager and Software Security roles often conduct the practices of the governance group, for which NIST 800-161 provides broad coverage of P-SSCRM tasks in this group, although it falls just short of providing 100% coverage of the P-SSCRM Governance group tasks. Fourteen of the 19 Product tasks came from the SSDF framework. Product tasks are done by the Architect/Developer role, and the SSDF is a development framework. Thirteen of the 23 Environment tasks came from the CNCF SSC framework. With its focus on the cloud environment, the CNCF SSC framework contains practices to protect the build infrastructure and computing environment. Finally, 5 of 8 tasks from the Deployment group come from SSDF. The purpose of including the OpenSSF Scorecard metrics in the table is to provide a longitudinal status of the ability of software ecosystems to automatically detect evidence that a task has been conducted on a software product/project. Currently, only 6 Product, 2 Environment, and 1 Deployment tasks can be automatically detected.

Table 1: Where did all the tasks come from? Number of tasks per group

| Framework | Governance | Product | Environment | Deployment | Total # tasks |
|---|---|---|---|---|---|
| P-SSCRM | 23 | 19 | 23 | 8 | 73 of 73 |
| EO 14028 | 12 | 17 | 6 | 7 | 42 of 42 |
| NIST 800-218 SSDF | 12 | **17** | 6 | **7** | 42 of 42 |
| Self-attestation | 9 | 11 | 5 | 4 | 29 of 29 |
| BSIMM13 | 17 | 14 | 2 | 4 | 37 of 125 |
| SLSA | 2 | 1 | 3 | 0 | 6 of 6 |
| NIST 800-161 | **20** | 10 | 9 | 5 | 44 of 183 |
| OWASP SCVS | 1 | 5 | 5 | 0 | 11 of 11 |
| S2C2F | 3 | 7 | 3 | 2 | 15 of 15 |
| CNCF SSC | 4 | 6 | **13** | 1 | 24 of 24 |
| OpenSSF Scorecard | 0 | 6 | 2 | 1 | 9 of 9[4] |

[4] OpenSSF Scorecard has 18 metrics. Fifteen (15) of these mapped to 9 tasks with five activities having more than one OpenSSF metric.

Software organizations that sell software to the US government or to another organization that sells software to the US government often associate software supply chain security with complying with EO 14028, which in turn means adopting all (32) and self-attesting to a subset (23) of the tasks of the SSDF. These tasks associated with EO 14028, the SSDF, and self-attestation are shown in lines 3-5 of Table 1. As the “D” in SSDF stands for Development, the SSDF is developer-focused, comprising tasks for the Developer/Architect and Software Security roles. *However, the SSDF is not enough*. Reducing software supply chain risk holistically involves other roles in the organization. The other six frameworks bring in the tasks of the other roles (Business manager, IT, DevOps) to secure the software supply chain and additional tasks for the Developer/Architect and Software Security roles.

P-SSCRM1 contains 37 tasks mapped to the BSIMM13; the scopes of P-SSCRM1 and BSIMM13 overlap in these 37 tasks. Using the P-SSCRM terminology, BSIMM13 quantifies the adoption of 125 tasks in establishing and nurturing a software security initiative in an organization and designing and building a secure product, including the inclusion of third-party software. P-SSCRM1 quantifies the adoption of tasks of protecting an organization from risks associated with the software supply chain. As shown in Table 1, each group in the P-SSCRM1 contains tasks from BSIMM13 but adds additional tasks, particularly in the Environment and Deployment groups commensurate with a focus of supply chain security on closing off the environment and build infrastructure attack vector.

P-SSCRM1 contains 44 of the 183 tasks in the NIST Cybersecurity Supply Chain Risk Management Practices for Systems and Organizations (800-161r1) which are specifically the tasks identified in NIST 800-161r1 as mapping to Executive Order 14028. The scope of NIST 800-161r1 includes manufacturing/hardware.

One Task, *Decommission assets* (E.1.6), was added that was not part of any of the ten frameworks. This task was added late in P-SSCRM development after several presentations of P-SSCRM and most of the interviews because of commentary that emerged in the conversations about the dangers of silent abandonment of a live system or product. These live systems can become an attractive attack vector because component updates and system monitoring may cease. Task E.1.6 involves decommission associated accounts, machines, data, keys, and passwords when a system goes end-of-life.

| Task Name | Objective | Definition | Question(s) | References |
|---|---|---|---|---|
| **GOVERNANCE (G):** Tasks that focus on the organization, measurement of a secure software supply chain, decision-making policies, accountability to third-party obligations, and compliance with legal and regulatory requirements. | | | | |
| **G.1 Perform compliance:** Compliance is following established guidelines or specifications, possibly demonstrated through an audit. | | | | |
| G.1.1 Organizational security requirements | Organizational security requirements, such as those imposed by standards and regulations, are included in the SDLC. | Identify, document, communicate, and maintain security requirements and policies for the organization's software development infrastructure and secure SDLC. Maintain the requirements and policies over time. Incorporate constraints imposed by standards and regulations and customer-driven security requirements. | Do you have a defined secure SDLC that the engineers are aware of? How do you define security requirements and policies for the organization, its development infrastructure, contributions, and processes? How are these requirements and contributions maintained over time? How are constraints imposed by regulatory and compliance drivers included in these requirements, policies, and the SDLC? | **EO:** 4e(ix)<br>**SSDF:** PO.1.1<br>**BSIMM:** CP1.1, CP1.2, CP1.3, SR1.1, SR2.2, SR3.3<br>**800-161:** SA-15<br>**CNCF SSC:** C: Establish and adhere to contribution policies<br>**Self-attestation:** 2 |
| G.1.2 Software license conflict | Software licenses that conflict with the organization's policies are identified. | Software licenses may or may not allow certain types of usage, contain distribution requirements or limitations, or require specific action if the software is modified. Risk is increased if the licenses of components conflict with an organization's policies. Software licenses should be documented and tracked to trace the users and use of licenses to access control information and processes according to software usage restrictions. License metadata should be recorded during the build and made available in the SBOM. | How do you scan check that the software licenses for tools or third party components comply with your organization's use policies? Is the process automated? Do you document and track users and uses of software licenses relative to access control policies and software usage restrictions? | **800-161:** CM-10<br>**OWASP SCVS:** 5.12<br>**S2C2F:** SCA-2<br>**CNCF SSC:** AU: Scan software for license implications |
| G.1.3 Produce attestation | Produce evidence of the use of secure software development practices. | Configure tools to generate artifacts to create an audit trail of the use of secure software development practices that conforms with record retention requirements and preserves the integrity of the findings and the confidentiality of the information. Assign responsibility for creating artifacts that tools cannot generate. Attestation should be immutable and published in the source repository releases, in the package registry, or elsewhere with their existence in a transparency log. | Can you tell me about automated or manual processes for producing artifacts that attest to the use of secure development practices? Where are the artifacts stored? What is your autitability/retention requirements for these artifacts? Is responsibility assigned for creating needed artifacts that tools cannot generate? Do you use a framework, like in-toto, to produce authenticated meta-data about artifacts, such as for attestation? Do you need to provide self-attestation for your product? Is the attestation immutable and published in the source repository releases, in the package registry, or elsewhere with their existence in a transparency log. | **EO:** 4e(i)(F), 4e(ii), 4e(v)<br>**SSDF:** PO.3.3<br>**BSIMM:** SM1.4, SR1,3<br>**800-161:** SA-15, AU-2, AU-3, AU-12<br>**SLSA:** Distributing attestation<br>**Self-attestation:** 1f |
| G.1.4 Deliver provenance | By providing provenance data, enable customers to analyze the integrity and verifiability of the actions performed while writing code, compiling, testing, and deploying software by making transparent the steps that were performed, by whom and in what order. | Automatically collect and provide provenance data for generated products. For provenance data, including how the artifact was built, including: what entity built the package, what build process they used, and the top-level inputs to the build, such that the data's authenticity and integrity can be verified and is generated automatically from the build service using a convention common in the package ecosystem. | Can you tell me more about your provenance data generation process (a) what authentications take place relative to the production of provenance data? (b) Is provenance data automatically generated/verified by the build service or does it involve a manual process or the ability for alteration or is it read-only? (c) is a signing key used for a digital signature? Stored in a secure management system accessible only by the build service? (d) do provenance records include the initial state of the machine, VM or container; include all user-specified build steps? | **EO:** 4e(vi), 4e(vii), 4e(x)<br>**SSDF:** PS.3.2<br>**SLSA:** Build L1: Prevenance exists<br>**800-161:** SR-4<br>**OWASP SCVS:** 6<br>**CNCF SSC:** V: Ensure clients can perform verification of artifacts and associated metadata<br>**Self-attestation:** 3 |
| G.1.5 Deliver SBOM | By providing SBOM, enable customers to analyze the contents of the final software package, including the version of the depedencies. | Generate (preferably during build time) and provide a SBOM (in a machine-readable, NTIA-supported format) for generated products. SBOMs should be digitally signed using a verifiable and trusted key. VEX data for all compone | Do you deliver an SBOM and provenance data that satisfy your contracts with software acquirers, such as in a standard-base format? How and when is the SBOM built? What tool do you use? Do you digitally sign the SBOMs? Do you produce a VEX supplement for your SBOM? What is your process for generating the VEX information? | **EO:** 4e(vi), 4e(vii), 4e(x)<br>**SSDF:** PS.3.2<br>**BSIMM:** SE3.6<br>**800-161:** SR-4<br>**OWASP SCVS:** 1.4, 6<br>**S2C2F:** REB-3, REB-4<br>**CNCF SSC:** SM: Generate immutable SBOM of code<br>**Self attestation:** 3 |
| **G.2 Develop security policies:** Establishing organizational roles and tasks for driving internal security standards in alignment with the business purpose of the organization. | | | | |
| G.2.1 Upper management support | Upper management understands the business risks of insecure software and supports the resources necessary for secure software development. | An upper management (e.g., c-suite) leadership team understands the business risk of insecure software development violating compliance and privacy obligations. The team is responsible for the entire software development process and deploying secure software to production. The commitment of this team is communicated to personnel associated with development-related roles and is backed up by the commitment to allocating adequate resources and making sometimes-difficult business decisions for security over faster release dates. | Can you share about upper management's commitment to secure software development? In what way is this commitment publicly demostrated and communicated to development roles and responsibilities? Comment on the adequacy of resources dedicated to secure software development? | **EO:** 4e(ix)<br>**SSDF:** PO.2.3<br>**BSIMM:** SM1.3, CP2.5 |
| G.2.2 Secure SDLC checks | Criteria throughout the SDLC are used to check the software's security during development. | Define criteria for a secure SDLC and associated software security checks that indicate how effectively software resulting from the SDLC meets the organization's expectations. These checks include key performance indicators (KPI), vulnerability severity scores, and security checks included in the "definition of done" in an agile process and may be used for go/no-go decisions. The use of automated tools aids thoroughness, objectivity, and efficiency of these checks. | What kind of criteria for security checks has been established, such as for security testing results? How can these checks indicate if security practices are being used and secure software is being developed? How are these checks tracked through the SDLC? Are the checks automated or manual? | **EO:** 4e(iv), 4e(v)<br>**SSDF:** PO.4.1<br>**BSIMM:** SM1.4, SM3.3<br>**800-161:** SA-15<br>**Self-attestation:** 4 |
| G.2.3 Roles and responsibilities | Ownership for security tasks throughout the SDLC at the organizational level and at the product team- and operational- level are established and visible. | Throughout the organization, create new roles and alter responsibilities for existing roles to incorporate security tasks and practices, as appropriate, into the SDLC and to keep management educated and informed on security issues. These roles can be centralized for the organization to promote thought leadership among developers and architects; and distributed throughout the organization in a network of security advocates. | What security roles and responsibilities have been created to encompass all parts of the SDLC - both within the development teams as well as at an organizational level? | **EO:** 4e(ix)<br>**SSDF:** PO.2.1<br>**BSIMM:** SM2.3, SM2.7, CR1.7<br>**800-161:** SA-3 |
| G.2.4 Security code review policy | Guidelines on which code should undergo a security-focused manual or automated review are communicated. | The policies for whether security-focused code review (a person looks directly at the code to find issues) and security-focused code analysis (tools are used to find issues in code, either in a fully automated way or in conjunction with a person) should be conducted based upon the characteristics/criticality of the software and its stage of development. Policies are established for both code developed in-house and third-party code. | What kind of guidelines are set in place for which code should be manually or automatically reviewed from a security perspective? How are these guidelines communicated and enforced? | **EO:** 4e(iv)<br>**SSDF:** PW.7.1<br>**BSIMM:** CR1.4, CR1.5<br>**800-161:** SA-11<br>**Self-attestation:** 2, 4 |

| Task Name | Objective | Definition | Question(s) | References |
|---|---|---|---|---|
| G.2.5 Asset inventory | Hardware and software assets are inventoried to enable incidence response; system analysis; traceability for critical components; and reliable identification when assets need to be changed or decommissioned. | Maintain a system component inventory, including hardware, software licenses, software versions, direct and transitive component owners, containers, machine names, and network addresses. Maintain an operations software inventory, including a map of source code; open source incorporated during the build and dynamic production; software deployments and related containerization, orchestration, and deployment automation code with respective owners. Unique identifiers for the inventoried assets should be established. Suppliers should also produce an asset inventory. Particular attention should be placed when a product or system is retired. | How is a system component inventory maintained for hardware, software licenses, software versions, component owners, machine names, and network addresses with identifiers for each of these? Is the components inventory process automated or manual? Are the artifacts machine readable (such as through SBOM generation)? What process is used for the asset inventory to be updated when a product or system goes to end-of-life? | **BSIMM:** CMVM2.3, SM3.1<br>**800-161:** CM-8, IA-4, PM-5<br>**OWASP SCVS:** 1<br>**S2C2F:** INV-1<br>**CNCF SSC:** SM: Track dependencies between OS components |
| G.2.6 Protection of information at rest | Protect the confidentiality and integrity of information at rest. | Data protection at rest aims to secure the confidentiality and integrity of data stored on any storage device or network. Mechanisms to achieve confidentiality and integrity protections include using encryption and file share scanning. Provisions for protecting information at rest should be included in agreements with suppliers, developers, system integrators, external system providers, and other service providers. | What provisions are made for the protection of information at rest included in agreements with suppliers, developers, system integrators, and external system providers? What policies are set in place for the use of cryptography in the organization's security policy? | **800-161:** SC-28 |
| **G.3 Manage suppliers:** Tasks to require that third-party suppliers employ adequate security measures to protect information, applications, and services provided to the organization. | | | | |
| G.3.1 Security-related contract terms | Component, system, and service acquisition policies include the inclusion of adherence to security policies, security requirements, and secure SDLC practices that are compatible with compliance requirements. | Compliance requirements, security requirements, and secure SDLC practices are included in vendor contracts with specified means of adherence enforcement. Examples include supplying an SBOM, self-attestation of security practices and provenance information; having a vulnerability disclosure program and incident response plan; and a security training program. The vendors include component, cloud, middleware providers, container and orchestration providers, and contractors. | What requirements language is included in system and services acquisition policies and contracts related to adherence to security policies, security requirements, and secure SDLC practices compatible with compliance requirements? Do you require SBOMs, provenance data, and attestation of adhering to security practices from your suppliers? How is adherence to contract terms verified? | **EO:** 4e(vi)<br>**SSDF:** PO.1.3<br>**S2C2F:** ING-1<br>**BSIMM:** CP2.4, CP3.2, SR2.5, SR3.2<br>**800-161:** SA-1, SA-4, SA-9, SR-3, SR-4, SR-5, SR-6<br>**Self-attestation:** 3 |
| G.3.2 Separation of duties | Reduce the potential for abuse of authorized privileges and the chance of collusion when acquiring components, systems, and services. | Ensure that appropriate separation of duties is established for decisions that require the acquisition and administration of information systems and the acquisition of components entering the supply chain. Separation of duties can be used to deny contracted developers the privilege to promote code they wrote from development to the production environment. Separation of duties can prevent collusion, for example, by ensuring personnel administering access control functions do not also administer acquisition. | How is separation of duties established for decisions that require the acquisition of information systems and supply chain components, such as components entering the enterprise's supply chain or contracted developers promoting code from development to production? | **800-161:** AC-5 |
| G.3.3 Information disclosure | Contract language requires that a vendor monitors for information disclosure and notifies the enterprise of information disclosure. | Unauthorized disclosure of information is a form of data leakage. Monitoring should be in place for contractor systems to detect the unauthorized disclosure of any data, and contract language should include a requirement that vendors notify of unauthorized disclosure of information. | What kind of contract language is in place to stipulate that contractor systems conduct monitoring to detect unauthorized disclosure of any data? What kind of contract language is in place to require that vendors notify of unauthorized disclosure? | **800-161:** AU-13 |
| G.3.4 Session audits | Identify security risks in the supply chain. | Include contract employees and prime contractors in session audits to identify security risks in the supply chain. Session audits can include monitoring keystrokes, tracking websites visited, and recording information or file transfers and may involve implementing specialized session capture technology. As such, the privacy risks of session audits should be considered, and session audits may only be activated under certain circumstances, e.g., the organization is suspicious of a specific individual. | Under what circumstances are contract employees included in audits to identify security risks in the supply chain? | **800-161:** AU-14 |
| G.3.5 Notification agreement | Timely notification of security threat and product end-of-life | Require suppliers to establish agreements and procedures for notification and monitoring capabilities, including notification of being the target of a supply chain threat. Timely notification of compromises and potential compromises in the supply chain is essential for an organization to initiate an incident response. Establish a minimum amount of time a vendor must declare that a product will be declared end-of-life and will no longer be supported and understand what end-of-life options exist (e.g. replace, upgrade, migrate to a new system, etc.). | What notification agreements and monitoring capabilities are established with your suppliers related to supply chain threats or incidents? How much notification does a vendor provide when a product goes end-of-life? What end-of-life options exist? | **800-161:** SR-8 |
| **G.4 Training:** Educating all personnel in role-specific information about the secure software development, including awareness, technical skills, and emergency response. | | | | |
| G.4.1 Role-based training | Provide security training for all personnel involved in software development | Provide literacy and role-based training on software security and secure software supply chain. This training should include information about the firm's security culture, the secure software development lifecycle, containerization and security orchestration, common security mistakes, technology such as CI/CD and DevSecOps, and recognizing insider threat. Require this training as part of onboarding and periodically. | Can you tell me about the training given to employees and to vendors related to software supply chain security? Do you require team members to regularly participate in secure software architecture, design, development, testing, and software supply chain training and monitor their completion of training? Is the training specific to their roles, development tools, and language? | **EO:** 4e(vi)<br>**SSDF:** PO.2.2<br>**BSIMM:** T1.1, T1.7, T1.8. T2.5, T2.8, T2.9, T3.1, T3.2<br>**800-161:** AT-2, AT-3, SA-16 |
| G.4.2 Contingency training | Provide training on procedures in the event of a security emergency. | Conduct contingency training to prepare for emergency response, backup operations, and post-disaster recovery to ensure the availability of critical information resources in emergencies. Critical suppliers should also be included in the training. | Can you tell me about contingency training for preparing for emergency response, backup operations, and post-disaster recovery? | **BSIMM:** T1.1<br>**800-161:** CP-3, IR-2 |
| G.4.3 Gather attack trends | Stay current on attack trends. | Have a process for continuously learning about emerging attack trends and vulnerabilities. Gather information from software acquirers, users, and public sources on potential vulnerabilities in the software and third-party components that the software uses, and investigate all credible reports. Use a subscription to a cyber threat intelligence feed, attend technical conferences, monitor attacker forums, and study trends within the enterprise. Make this information on attack trends actionable and useful for developers, testers, security operations, and others to identify vulnerabilities in existing products, perform improved threat modeling and security architecture, and evolve the SDLC. | How do you monitor new attack trends and vulnerabilities applicable to your software, such as by monitoring a cyberthreat intelligence feed and attacker forums, or attending conferences? | **EO:** 4e(iv), 4e(vi), 4e(viii), 4e(ix)<br>**SSDF:** RV.1.1<br>**BSIMM:** AM1.5, CMVM1.2<br>**800-161:** SI-4, SI-5<br>**Self-attestation:** 2 |
| **G.5 Assess and manage risk:** proactively analyzing, mitigating, and managing software supply chain risk and achieving the objectives of a software security program | | | | |
| G.5.1 Criticality analysis | Identify critical system components and functions by performing a criticality analysis | Perform a criticality analysis of system components, functions, or services to assign cybersecurity supply chain risk management activities commensurate with the analysis based upon the likelihood and impact of an attack. Not all system components, functions, or services necessarily require significant protection. Items to consider in the analysis include system assets/data involved in the product, applicable laws, regulations, policies, standards, system functionality requirements, system and component interfaces, and system and component dependencies. The criticality analysis impacts the procedures contractually imposed on vendors. | What process do you have for performing criticality analysis as input to assessments of supply chain risk management activities? | **BSIMM:** AA1.4<br>**800-161:** RA-9 |
| G.5.2 Track security risks and decisions | Record and track the software's security risk-based exceptions and mitigation plans | Record when a design tradeoff, vulnerability decision/exception, or component choice has been made that incurs security risk and a mitigation plan to reduce that risk including with vendors. Periodically re-evalute these exceptions and mitigation plans, potentially including a review board that can provide security guidance in design guidance, patterns, standards, features, and frameworks. | How are the responses to security risks and design decisions recorded, including how mitigations are to be acheived? How are approved exceptions to the security requirements periodically evaluated? | **EO:** 4e(v)<br>**SSDF:** PW.1.2<br>**BSIMM:** SFD3.1, SM3.5 |

| Task Name | Objective | Definition | Question(s) | References |
|---|---|---|---|---|
| G.5.3 Security metrics | Provide the basis for the measurement of an effective plan for tracking and realizing software security objectives within an organization. | Identify and regularly review metrics to measure the effectiveness of the software security program, including establishing Key Performance Indicators (KPIs). The regular review can drive budgeting and resource allocations and measure performance against risk appetite and risk tolerance statements. Publish data/dashboards about the state of software security within the organization Automate metrics collection via security telemetry, as possible, to gather measurements to enhance efficiency and objectivity. | What security metrics do you track to indicate how you are doing relative to developing secure software products and having an effective and efficient software security program (e.g. security outcomes, MTTR)? Are these metrics collected automatically or manually? | **BSIMM:** SM2.1, SM3.3 |
| G.5.4 Data-informed product decisions | Make security decisions on software release based upon criteria for checking the security of the software. | Implement processes and mechanisms, automated when possible, to gather and safeguard information in support of security decison-making at the product level. This information can be used to drive SDLC change at the product or organizational level and/or drive a security risk exception process. | Do you use the toolchain to gather information that informs security decision-making automatically? If not, is this information collected and reviewed manually? | **EO:** 4e(iv), 4e(v)<br>**SSDF:** PO.4.2<br>**BSIMM:** SM1.4, SM2.2<br>**800-161:** SA-15<br>**Self-attestation:** 4 |
| **PRODUCT (P):** Tasks to lead to deploying a secure product with minimal vulnerabilities with associated required attestations and artifacts. | | | | |
| **P.1 Develop security requirements:** the development of software-related objectives and expectations to protect the service and data at the core of the application. | | | | |
| P.1.1 Product security requirements | Identify and document security requirements for organization-developed software | Identify and document security requirements for organization-developed software, including risk-reducing software architecture and design choices, security patterns, and translating compliance constraints to requirements. Examples include using memory-safe languages and secure frameworks isolation and sandboxing component strategies; code modularity; security features; secure-by-design components; application containers; and product features that aid in secure deployment, operation, and maintenance. Containers can be used as a strategy for tighter coupling of an application and its dependencies, immutability, and some isolation benefits. Maintain these requirements over time. | How are risk-reducing security architecture and design requirements for products considered and developed? Which of the following strategies are considered: the use of memory-safe languages, secure frameworks, isolation and sandboxing? Are completion and adherence tracked? | **EO:** 4e(iv)<br>**SSDF:** PO.1.2<br>**BSIMM:** CP1.1, CP1.2, CP1.3, CP2.1, SE2.5, SFD1.1, SFD2.1, SDF3.2, SR1.3<br>**800-161:** SA-8 |
| P.1.2 Software release integrity | Provide software acquirers assurance that the software they acquire is legitimate and has not been tampered with. | Use code protection mechanisms, such as the use of an established certificate authority for code signing, to enable the attestation of the provenance, integrity, and authorization of important code. Make software integrity verification information, such as cryptographic hashes for release files, available to software acquirers on a well-secure website. | What software integrity artifacts do you make available to software acquirers, such as cryptographic hashes of release files? Do you use a certificate authority for code/commit signing? How often do you review code signing processes, such as certificate renewal and key rotation? Do you ensure use a dedicated, protected signing server when signing is required? What kind of license is declared for a project? | **EO:** 4e(iii)<br>**SSDF:** PS.2.1<br>**BSIMM:** SE2.4<br>**OWASP SCVS:** 6<br>**OSSF Scorecard:** signed-release, licenses<br>**S2C2F:** REB-2<br>**Self-attestation:** 2 |
| **P.2 Build security in:** use software development practices and processes that will lead to the development of secure software products | | | | |
| P.2.1 Security design review | Decrease the number of design flaws and security vulnerabilities introduced during the architecture and design phases | Enumerate possible threat vectors. Conduct threat modeling and attack surface analysis to identify weaknesses in the software architecture and design such that the system is not resistent to attack. Identify missing security features and requirements. Identify unused components in the design. Address the identified security risks through re-design, or develop and track a mitigation plan for exceptions. | For all critical software components and external services that your team operates and owns, has a qualified person not involved in the design reviewed the design and conducted an attack surface analysis and threat model? What kind of approaches are you to narrow attack vectors? What kind of tool or a methodology (like STRIDE) do you use to structure your threat modeling? Do you perform analysis such that only required modules are included in the product and unused modules are uninstalled and removed, decreasing the attack surface? How are unused modules identified - such as "debloating" a product for unused components and containers? | **EO:** 4e(iv), 4e(v), 4e(ix)<br>**SSDF:** PW.1.1, PW.2.1<br>**BSIMM:** AA1.1, AA1.2, AA2.1, AA3.1<br>**Self-attestation:** 4 |
| P.2.2 Secure coding | Decrease the number of security vulnerabilities introduced during source code creation | Follow secure coding standards and practices appropriate to the development languages, APIs, and environment. Examples include validating all inputs, encoding outputs, and avoiding unsafe functions. Use development environments and static analysis tools that support compliance with secure coding practices as source code is being implemented. | What kind of secure coding practices are used? How are these practices communicated to developers and enforced? | **EO:** 4e(iv)<br>**SSDF:** PW.5.1<br>**BSIMM:** SR3.3, CR1.4, CR3.5<br>**Self-attestation:** 4 |
| P.2.3 Secure-by-default implementation | Improve the security of software at the time of installation | Reduce the likelihood of software deploying with weak security settings by defining secure deployment parameters so that the default settings are secure and do not weaken the security functions provided by the platform, network infrastructure, or services. | Are the default installation settings for products, platform, network infrastructure, services defined to be secure by default? | **EO:** 4e(iv)<br>**SSDF:** PW.9.1 , PW.9.2<br>**BSIMM:** SE2.2<br>**800-161:** SA-5<br>**Self-attestation:** 4 |
| P.2.4 Standard security features | Reduce introducing new vulnerabilities by reusing standardized and proven security features | Build support for standardized rather than proprietary security features, such as using existing log management, identity management, access control, or vulnerability management systems. These reused components are more likely to have their security posture already checked. | Can you tell me about your philosophy of whether it is better to use available security features or to "roll your own," for example, log management, identity management, access control, and vulnerability management? Or a combination of in-house/proprietary and standard features? What factors determine if you use in-house or standard features? | **EO:** 4e(ix)<br>**SSDF:** PW.1.3<br>**BSIMM:** SFD1.1, SFD3.2 |
| P.2.5 In-house components | Maintain components built in-house | Well-secured in-house components and scripts are built following a secure SDLC process when third-party components cannot meet development needs. Similar to the processes for third-party components, in-house components should be kept in a repository, maintained, and regularly scanned for vulnerabilities, with new versions deployed through the organization as needed. | What processes are used to scan in-house components built using SDLC processes for vulnerabilities and how is the process similar or different to what is done for third-party components? Are new versions deployed throughout the organization as needed? What causes a need for a new version? Are in-house components monitored to ensure they are regularly maintained? | **EO:** 4e(ix)<br>**SSDF:** PW.4.2<br>**BSIMM:** SFD2.1 |
| **P.3 Manage component and container choices:** software supply chain risk can be reduced by careful choice and handling of third-party components and containers | | | | |
| P.3.1 Component and container choice | Make informed third-party component and container choices | Consider component characteristics, such as the use of secure SDLC practice, evidence of ongoing maintenance, and open vulnerabilities, as quality indicators of direct and transitive dependencies. OpenSSF Scorecard metrics may be used as an indication of the security posture of a component. Components with the least functionality reduce the attack surface. Have a deny-list that prevents malicious components from being consumed. | What kind of approval process for you have, if any, for third-party libraries and containers included in a product? What do you consider when selecting a third-party component or container? Do you consider solutions with the least functionality to reduce the attack surface? How would the approval process handle dependencies that are no longer receiving updates? Do you look at the OpenSSF Scorecard metrics for a component? Are you concerned when a component contains binary artifacts that cannot be easily reviewed and more easily sneak in a vulnerability? Do you have a Deny List to prevent the choice of a vulnerable and malicious components? | **BSIMM:** SR2.4<br>**800-161:** CM-7<br>**OSSF Scorecard:** Binary-artifacts; contributors<br>**S2C2F:** ING-3<br>**CNCF SSC:** SM: Second and third-party risk management |

| Task Name | Objective | Definition | Question(s) | References |
|---|---|---|---|---|
| P.3.2 Trusted repositories | Obtain candidate packages and containers from trusted ecosystems or rebuild | Trusted public repositories may require signed packages and provide the means to verify the signatures; the packages and containers from these ecosystems should still be scanned. Organizations should host scanned repositories for high-assurance software and restrict build machines to only those sources. | Do you search for candidate components in package managers trusted by your organization? Do you use container registries you trust? | **OWASP SCVS:** 1.2<br>**S2C2F:** ING-1<br>**CNCF SSC:** V: Define and prioritize trusted package managers |
| P.3.3 Require signed commits | Utilize legitimate components software that has not been tampered with. | Using code that has been contributed with signed source code commits provides an integrity mechanism. | Do you require the software producers sign code commits of the components and tools you bring into your product? | **BSIMM:** SE2.4<br>**CNCF SSC:** V: Require signed commits. |
| P.3.4 Vetted third-party component and container repositories | Engineers can choose from organization-approved components | Components, binaries, and containers chosen from public ecosystems undergo a vetting process, including reviewing component meta-data and provenance data; secure composition analysis; binary composition analysis; and other security scanning. Rebuild software dependencies from source code when possible to remove additional software that may be contained in a compiled version. Establish repositories to host organization-approved components and containers. | Do you have a local component repository in which binaries and source code of third-party components are stored? Do you rebuild a component from source when possible? | **EO:** 4e(iii), 4e(vi), 4e(x)<br>**SSDF:** PW.4.1<br>**BSIMM:** SFD2.1, SFD3.2, SR2.4, SR2.7<br>**SLSA:** Build L1: Prevenance exists<br>**800-161:** SR-3, SR-4, SR-11<br>**OWASP SCVS:** 4.1, 5<br>**S2C2F:** SCA-1, SCA-4, SCA-5, ING-2, ING-4, ENF-2<br>**CNCF SSC:** SM: Verify 3rd party artifacts and open-source libraries; SM: Build libraries based upon source code<br>**Self-attestation:** 2, 3 |
| P.3.5 Prevent component vetting bypass | Ensure developers are not bypassing the component vetting process | Audit to ensure developers are consuming components that have gone through the approved vetting process even on their own machines because developer endpoints can become infected. Restrict product build to only packages in the repository of vetted components. | What mechanisms are in place, if any, to audit that developers are consuming third party code through the approved ingestion method and not bypassing the vetting process? | **S2C2F:** AUD-2 |
| **P.4 Discover vulnerabilities:** Use automated and manual vulnerability discovery techniques to identify previously-undisclosed vulnerabilities in in-house and third-party code | | | | |
| P.4.1 Security code review | Detect security vulnerabilities introduced during architecture, design, and source code creation including those injected with malicious intent | Perform peer code review on in-house developed code, AI-generated code, and some third-party code based on the organization's secure coding standards. Create and use review checklists that may be informed by "Top N" vulnerability lists. While the manual review of another programmer before check-in is essential for finding vulnerabilities injected with malicious intent, supplementing code review with automated tools, such as static analysis tools, is beneficial. Record and triage discovered issues and recommended remediations in the development team's workflow or issue-tracking system. | How does the team perform security-related code reviews? Is a secure coding standard used? How do discovered vulnerabilities get recorded, tracked, and remediated? | **EO:** 4e(iv), 4e(v)<br>**SSDF:** PW.7.2<br>**BSIMM:** CR1.2, CR1.4, CR1.6, CR2.6, CR2.7, CR3.4, CR3.5<br>**OSSF Scorecard:** Code review<br>**CNCF SSC:** C: Enforce an independent four-eyes principle<br>**Self-attestation:** 4 |
| P.4.2 Automated security scanning tools | Choose tools to efficiently discover previously-undetected vulnerabilities | Specify which tools or tool types must or should be included in each toolchain to reduce human effort in discovering vulnerabilities and mitigating identified security risks and to improve the accuracy, reproducibility, usability, and comprehensiveness of security practices in the SDLC. Tools commonly include SAST, DAST, IAST, SCA, and secret detection tools. Specify how the toolchain components will be integrated, such as through the CI/CD pipeline. Follow recommended practices for tool selection/scanning, deploying, operating, and maintaining tools and toolchains. | What security tools are in your toolchain? SAST? DAST? IAST? How often is the scanning done for third-party components? Do you use a tool to detect secrets in code? What happens with the reported vulnerabilities? Do you scan your tools for vulnerabilities? | **EO:** 4e(i)(F), 4e(ii), 4e(iii), 4e(v), 4e(vi)<br>**SSDF:** PO.3.1; PO.3.2<br>**BSIMM:** CR1.4, ST1.4, ST2.5<br>**800-161:** SA-15<br>**OSSF Scorecard:** SAST; Fuzzing<br>**CNCF SSC:** A: Prevent commiting secrets; A: Automate software security scanning; AU: Scan software for vulnerabilities<br>**Self-attestation:** 1f, 2, 3 |
| P.4.3 Automated vulnerability detection | Detect vulnerabilities before deployment through the use of automated tools to reduce the window of opportunity for attackers | Review, analyze, and/or test the software's code to identify or confirm the presence, remediation, and disclosure of previously undetected vulnerabilities. Record and triage discovered issues and recommended remediations in the development team's workflow or issue-tracking system. | How is code reviewed, analyzed, and tested to identify or confirm the presence, remediation, and disclosure of previously-undetected vulnerabilities such as through automated tools? | **EO:** 4e(iv), 4e(vi), 4e(viii)<br>**SSDF:** RV.1.2<br>**BSIMM:** CMVM3.1<br>**800-161:** SA-11, RA-9, SI-7<br>**OWASP SCVS:** 6<br>**Self-attestation:** 3, 4 |
| P.4.4 Executable security testing | Discover vulnerabilities only detected through executable testing | Executable testing, such as done by in-house testers or third-party penetration testers, is performed to find previously-undiscovered vulnerabilities. In-house testers should be guided by design review results, threat modeling, security requirements, and the security mechanisms in security features. Testing should include containers. Automate security tests as possible. Scope the testing, design the tests, perform the testing, and document the results, including recording and triaging all discovered issues and recommended remediations in the development team's workflow or issue tracking system. Test the build process and scripts. | How are security features tested? How are test cases developed to test the security of an application? Is security testing factored into a product's test suite? Is executable code testing, such as in-house or external penetration testing, to find vulnerabilities not found by previous reviews, analysis, and testing? Do you do executable testing on containers? Do you test your build process, scripts, GitHub actions? | **EO:** 4e(iv), 4e(v)<br>**SSDF:** PW.8.1, PW.8.2<br>**BSIMM:** ST1.1, ST1.3, ST1.4, ST2.4, ST2.5, ST2.6, ST3.3, ST3.4, ST3.5, ST3.6, PT1.1, PT1.2, PT1.3, PT2.3, PT3.1, CMVM3.4<br>**800-161:** SA-11<br>**OSSF Scorecard:** CI-Tests<br>**Self-attestation:** 2, 4 |
| P.4.5 Regular third-party compliance | Identify increased security risk of third-party and open-source components | Regularly check that third-party software components comply with the contractual requirements, such as following a secure SDLC, and fix detected vulnerabilities. Regularly scan components and containers to monitor for vulnerabilities and for evidence that artifacts are being maintained and have not been abandoned or deprecated. | How do you verify that third-party software components continue to comply with the requirements defined by the organization, such as through using a secure SDLC and the delivery of attestation of security practices? Do you check that components are being maintained and are not at end-of-life or have been abandoned? How are known critical vulnerabilities in third-party and open-source components and containers mitigated? | **EO:** 4e(iii), 4e(iv), 4e(vi), 4e(x)<br>**SSDF:** PW.4.4<br>**800-161:** SA-4, SA-9, SA-11, SA-15, SR-3<br>**OWASP SCVS:** 5<br>**S2C2F:** SCA-1, SCA-3<br>**Self-attestation:** 2, 3, 4 |
| **P.5 Manage vulnerable components and containers:** develop and implement a strategy for patching/upgrading components and containers to the latest secure version. | | | | |
| P.5.1 SBOM consumption | Utilize SBOM information to react to security incidents and to identify which components need to be updated or patched. | Obtain or generate an SBOM for a product that provides a clear and direct link to the dependencies and their versions used within a product. Ideally, the SBOM will provide signed metadata from the build process. Tools can automate analysis and obtain desired information from the SBOM, such as aligning with vulnerability data to identify vulnerabilities within the exact package contents. | Does anyone check the SBOM information about pending security vulnerabilities? How do you consume the SBOMs for your components? Do you have any specific experience being aided by information provided by the SBOM? How do you feel about the assistance provided by tools to help you consume an SBOM? If not now, do you have plans to consume SBOMs in the future? Do you consider and trust the VEX information in the SBOM? What expectations do you put on the component producers for the vulnerabilities identified in the SBOM? | **800-161:** SR-4<br>**CNCF SSC:** SM: Require SBOM from third-party supplier; SM: Track dependencies between OS components |
| P.5.2 Dependency update | Update vulnerable dependency when a fixed version is available | SCA tools, SBOM tools, and ecosystem tools (such as Dependabot) inform a product of vulnerabilities and provide an automated pull request for new versions of direct and transitive dependencies discovered vulnerable. Organizations need a process/strategy for updating dependencies - which may be manual or automated. Specify the software assets that require automated updates, defined from criticality/risk-based analysis. Rebuild, do not patch, containers with vulnerabilities | What is your strategy for updating dependencies based on SCA, SBOM, or automated pull requests? Do you have a strategy for updating all, such as through a tool such as Dependabot that automated PR? Does the project use tools to help update its dependencies e.g. Dependabot, RenovateBot? Are the software assets that require automated updates defined from a criticality/risk-based perspective? How are containers updated when a vulnerability is detected? | **EO:** 4e(viii)<br>**SSDF:** RV.1.1<br>**800-161:** SI-2<br>**OSSF Scorecard:** Dependency update tool<br>**S2C2F:** UPD-1, UPD-2, UPD-3<br>**Self-attestation:** 2, 3, 4 |
| **ENVIRONMENT (E):** Tasks to protect the confidentiality and integrity of source code, software components, and the build infrastructure from tampering and unauthorized access. | | | | |
| **E.1 Safeguard artifact integrity:** protect from unauthorized or accidental access and alteration of project artifacts | | | | |

| Task Name | Objective | Definition | Question(s) | References |
|---|---|---|---|---|
| E.1.1 Safely store release artifacts | Preserve release artifacts to help in the identification and analysis of vulnerabilities discovered after release. | Securely archive the necessary files and supporting data (e.g., integrity verification information, provenance data, metadata) to be retained for each software release. | What files and supporting data (like integrity validation, provenance, and metadata) do you securely archive and retain for each software release? | **EO:** 4e(iii), 4e(vi), 4e(x)<br>**SSDF:** PS.3.1<br>**OWASP SCVS:** 6.3<br>**Self-attestation:** 2, 3 |
| E.1.2 Version control | Prevent unauthorized charges to artifacts, both inadvertent and intentional. | Store project artifacts, including but not limited to source code, executable code, infrastructure as code, and configuration-as-code, in a repository with restricted access using the principle of least privilege based on the nature of the artifact. Use version control to track and store all changes to this code with accountability to an authenticated individual account with access granted to personnel, tools, and services. As appropriate, sign or encrypt artifacts. Indefinitely retain change history. | Are code and other important project artifacts stored with version control with access granted to strongly-authenticated personnel using the principle of least privilege? | **EO:** 4e(i)C, 4e(iii), 4e(iv)<br>**SSDF:** PS.1.1<br>**188-161:** SA-8, SA-10<br>**Self-attestation:** 2, 4 |
| E.1.3 Multi-factor authentication (MFA) | Decrease the chances an account will be compromised | Enforce multi-factor authentication (MFA) at the source code repository level by requiring a soft or physical token in addition to traditional passwords/credentials that are more likely to be compromised using techniques such as brute force guessing. | Is MFA mandated and enforced at the source code repository level? Are soft or physical tokens used? Is MFA enforced for endpoints? | **CNCF SSC:** SA: Enforce MFA |
| E.1.4 Developer SSH key | Decrease the chances developer account will be compromised | Use SSH keys or SSH certificates for developers and agents in CI/CD pipelines to access source code repositories from their development tools rather than traditional passwords/credentials that are more likely to be compromised. Passphrases should be used to protect SSH keys. | Do you avoid using passwords through use of SSH keys, SSH certificates, or short-term rotated Access Tokens? | **CNCF SSC:** SA: Use SSH keys |
| E.1.5 Branch protection | Provide a formal approval process for code changes to enforce adherence to software development processes and policies before code is introduced into a CI/CD system. | Branch protection settings to enforce security policies, such as requiring reviews, passing checks, preventing overwriting of history, and signed commits before code is accepted into the main branch. Branch protection can prevent forced pushings, overwriting history, and obfuscation of code changes. In high-risk, high-assurance environments, attribution of code changes, commit signing, and full attestation can be used with branch protection to prevent and detect complex attacks. | Do you use branch protection settings to enforce security policies, such as requiring reviews, passing checks, preventing overwriting of history, signed commits before code is accepted into the main branch? In high assurance, high risk enviroments, is commit signing with full-attestation (a.k.a the signature of the final commit once all signatures are verified) required? | **OSSF Scorecard:** Branch protection<br>**CNCF SSC:** V: Enforce full attestation and verification for protected branches; C: Use branch protection rules; C: Establish and adhere to contribution policies |
| E.1.6 Decommission assets | Prevent security attacks through live end-of-life systems and products | Live but abandoned systems may not get new security patches or be monitored for malicous activity, even though they may be vulnerable to new security threats and can be an attack vector. When a product or system has been declared end-of-life, decommission associated accounts, machines, data, keys, and passwords. | What decommissioning procedures are in place when a live product or system goes to end-of-life? | |
| **E.2 Safeguard build integrity:** protect from and detect malicious infiltration into software build infrastructure that could lead to the build and deployment of compromised products | | | | |
| E.2.1 Release policy verification | Ensure the products, materials, and processes used during the build pipeline adhere to the established product and organizational release policy | A release policy should be maintained as a policy template and outline the required workflow while the software is developed, built, tested, and packaged to ensure the integrity, authentication, and auditability of a software product from initiation to end-user installation. A framework, such as in-toto, can produce meta-data during the build pipeline to enable attestation to the steps in the desired workflow. The meta-data can be analyzed to ascertain whether the steps in the workflow have occurred and to produce cryptographic guarantees by hashing and signing the inputs and outputs of steps in the pipeline. | Do you have a templated build policy that specifies the required build workflow to ensure the integrity, authentication, and auditability? Do you verify the build policy in a cryptographically provable way, such as with in-toto? | **CNCF SSC:** BV: Cryptographically guaranteed policy adherence |
| E.2.2 Verify dependencies and environment | Ensure the build environment's sources and dependencies come from a secure, trusted source of truth | Analyze the integrity of build tools, components, and containers before bringing them into the pipeline. Validate the point of origin/provenance, checksums, and signatures in the downloading and ingesting processes. Request SBOM, provenance, and self-attestation to aid in achieving this goal. | Can you track a package back to its repo? Do you verify the integrity (i.e. digital signature or hash match) and the provenance data of each component and container? Do you validate that the build environment's sources and dependencies come from a secure, trusted source of truth? | **SLSA:** Build L1: Prevenance exists<br>**OWASP SCVS:** 1.10, 6.1<br>**S2C2F:** AUD-1, AUD-3, AUD-4<br>**CNCF SSC:** V: Verify third-party artifacts; BV: Validate environments and dependencies |
| E.2.3 Defensive compilation and build | Reduce vulnerabilities during compilation and build. | Determine which compiler, interpreter, and build tool features should be used to reduce vulnerabilities, such as producing compiler warnings for vulnerable code that are treated as errors; application of obfuscation techniques; approved configurations being used; verification of sources and manifests of dependencies; and approved tools configurations are available as configuration-as-code. | Do you use compiler, interpreter, and build tool features to detect vulnerabilities? Do you have a Deny List to prevent using a vulnerable and/or malicious component in the build? | **EO:** 4e(iv)<br>**SSDF:** PW.6.1<br>**SLSA:** 4_Reproducible build<br>**OWASP SCVS:** 3<br>**S2C2F:** REB-1<br>**CNCF SSC:** BV: Validate build artefacts through verifiably reproducible builds.<br>**Self-attestation:** 4 |
| E.2.4 CI/CD hosting and automation | Through automated build and protected environments, reduce human error and malicious actions and artifacts that cause the output of the build process to contain security vulnerabilities. | All CI/CD pipeline build steps should be automated with automation standardized across the enterprise via clear templated build pipelines that meet organizational standards. This automation should be treated as "pipeline as code" with modifications being reviewed and should be immutable. The build should run in an ephemeral and hermetic (isolated and sealed) environment with no parameters or network connectivity except to the hardened local repository of source code and dependencies and code signing infrastructure. The build runs on a hosted build platform that generates and signs the provenance. | Are your build steps defined in a build script, with the only manual step being to invoke the script? Is the build definition executed by the build service stored in a version control system fetched through a trusted channel with a trustworthy provenance chain? Does the build run on a hosted build platform? Is the build service run in an ephemeral environment, such as a container or VM, provisioned solely for the build? Is the build service run in an isolated environment free of influence from other build instances? Can the build output be affected by user parameters other than the build entry point? Are the build steps, sources, and dependencies declared up front with immutable references/use immutable artifacts stored in a local repository manager? Is the build run with no network access (i.e. hermetic)?<br>Are infrastructure, build scripts and GitHub Actions handled "as code", such as with code review, scanning, testing, and version control? Are configuration changes reviewed?<br><br>Are CI/CD pipelines standardized across the enterprise? | **EO:** 4e(iv)<br>**SSDF:** PW.6.2<br>**BSIMM:** SE2.4, SE3.2<br>**SLSA:** Build L2: Hosted build platform; Build L2: Hardened/isolated builds<br>**800-161:** SA-15, SR-9<br>**OWASP SCVS:** 3<br>**OSSF Scorecard:** Dangerous workflow; token permissions; pinned-dependencies<br>**CNCF SSC:** BV: Build and related CI/CD should be automated; BV: Standardize pipelines; BV: Build workers should be single-use;CE: Ensure software factory had minimal network connectivity; CE: Segregate the duties of each build worker; CE: Pass in build worker environment and commands; SA: Only allow pipeline modifications through "pipeline as code"<br>**Self-attestation:** 4 |
| E.2.5 Secured orchestration platform | Ensure each deployed workload meets predetermined security requirements | The build pipeline should be a series of hardened build steps implemented through a hardened container image stored within a secured repository and deployed through a hardened orchestration platform, such as Kubernetes. Orchestration processes can take advantage of built-in and added-on security features, such as checking for secrets and rollbacks to ensure that each deployed workload meets predetermined security requirements. | Are containers used throughout the build process? Do you deploy Kubernetes as the orchestration layer? | **BSIMM:** SE2.7<br>**CNCF SSC:** BA: Provision a secured orchestration platform |
| E.2.6 Reproducible builds | Provide a mechanism to confirm that no malicious backdoor injections have taken place during the build process | Follow recommended security practices to deploy, operate, and maintain tools and toolchains, including the reproducible build steps whereby identical input artifacts are rebuilt from source in a trusted build environment, non-determinism is eliminated, and the results can be cryptographically attested to be the same output in a bit-for-bit comparison. | Do you practice reproducible builds on different protected environments and compare bit-for-bit results? | **OWASP SCVS:** 3<br>**S2C2F:** REB-1<br>**CNCF SSC:** BV: Validate build artefacts through verifiably reproducible builds. |

| Task Name | Objective | Definition | Question(s) | References |
|---|---|---|---|---|
| E.2.7 Build output | Protect the integrity of build output | Write the output from the build process to separate storage from the inputs. A process separate from the build process should upload the artifact to the appropriate deployable repository. | Do you write artifacts from the build to a separate shared storage from the inputs? | **CNCF SSC:** CD: Write output to a separate secured storage repo |
| **E.3 Secure software development environment:** protect the software development environment from internal and external threats that can lead to compromise | | | | |
| E.3.1 Authentication | Enable authorization, traceability, and non-repudiation | Authenticate employees and contractors to enable access control, traceability, non-repudiation, chain-of-custody, and provenance for systems, software and services. | Are employees and contractors authenticated to enable access control, traceability, non-repudiation, chain of custody, and provenance for systems, software, and services? Where do you store access credentials (e.g. hashes for passwords) and secrets in a secured (e.g. encrypted) location, such as a secure vault? | **800-161:** IA-5, IA-9 |
| E.3.2 Environmental separation | Separate and protect each environment involved in software development (i.e. development, build, test, deployment) | Using network segmentation and access control to separate and protect each environment (e.g., endpoints, development, build, test, deployment) involved in software development to protect from internal and external threats. Remote access to the build environment should be rare, logged, and require multi-party approval. Log and monitor access, particularly priviledged access. Configure the environment's hosting infrastructure following a zero-trust architecture. | Are environments involved in software development (e.g. endpoints, development, build, test, deployment) separated and protected from physical and remote access to the build environment? Is access logged? What kind of approval is needed for access to the build environment? | **EO:** 4e(i)(A), 4e(i)(B), 4e(i)(C), 4e(i)(D), 4e(i)(E), 4e(i)(F), 4e(ii), 4e(iii), 4e(v), 4e(vi)<br>**SSDF:** PO.5.1, PO.5.2<br>**800-161:** SA-8, SA-15<br>**Self-attestation:** 1a, 1b, 1c, 1d, 1e, 1f, 2, 3 |
| E.3.3 Role-based access control | Controlling access to resources where permitted actions on resources are identified with roles rather than with individual subject identities; provide traceability between actors and actions | Define a set of roles and accounts with an associated access control/authorization level considering remote access requirements, the period for which access is needed, and careful vetting of contractor personnel. These roles can be used for access control to ensure proper traceability of actions and actors. A process should be implemented to manage mission-critical systems' temporary or emergency access. Role-based access is enforced by physical and logical access enforcement mechanisms. | How are authorization levels for accounts associated with roles and associated level of access control, including remote access, proper vetting of contractors, and that authorization does not exceed the period of performance? How are access control violations handled? | **800-161:** AC-2, AC-3, AC-6, AC-17, IA-2<br>**CNCF SSC:** A: Define individuals/teams responsible for code; C: Define roles aligned to functional responsibilities; SA: Define user roles |
| E.3.4 Information flow enforcement | Limit the information flow across trust boundaries to participants in the supply chain | Specify how information flow across trust boundaries is enforced to ensure only the required information is communicated to participants in the supply chain, including but not limited to suppliers, developers, system integrators, and external system providers. Requirements for transmission confidentiality and integrity are integrated into agreements with suppliers. The degree of protection should be based on the sensitivity of the information. Security mechanisms such as authentication, authorization, and encryption can be used to achieve enterprise confidentiality and integrity requirements. | How is information flow enforced to ensure only the required information is communicated across trust boundaries to participants in the supply chain through mechanisms that ensure transmission confidentiality and integrity? What kind of requirements for transmission confidentiality and integrity are integrated into agreements with suppliers? | **800-161:** AC-4, SC-8 |
| E.3.5 Baseline configuration | Provide the starting point for tracking changes to components, code, and settings throughout the SDLC | Establish a baseline configuration of the information system and the development environment, including documenting, formally reviewing, and securing the agreement of stakeholders. The baseline provides a starting point for tracking changes to components, code, and settings for traceability and provenance. | Is a baseline configuration of the information system and development environment in place to provide a starting point for tracking changes to components, code, and/or settings for traceability and provenance? Who sets the baseline? | **800-161:** CM-2 |
| E.3.6 Monitor changes to configuration settings | Prevent the tapering of information systems and networks through the monitoring of changes to configuration settings. | Monitor and audit configuration settings and change controls within the information systems and networks throughout the SDLC. Changes should be tested and approved before being implemented. Configuration settings should be monitored so designated employees can be alerted when a change has occurred. | What kind of monitoring and auditing of changes configuration settings is in place? What designated employees are alerted when an unauthorized change occurs? | **800-161:** CM-3, CM-6 |
| E.3.7 Boundary protection | Monitors and controls communications at the external boundary of the system and at key internal boundaries within the system | The enterprise should consider the trust boundaries and provide separation and isolation of development, test, and security assessment tools and external networks or information systems. The connections should only be through managed interfaces consisting of boundary protection devices arranged in accordance with an organizational security architecture. | How are development, test, and security assessment tools and operational environments separated and isolated from external networks or information systems and each other, as appropriate considering trust boundaries? How are trust boundaries considered, and how is the separation/isolation implemented? | **800-161:** SC-7 |
| E.3.8 Key rotation | Limit the impact if a key is compromised. | Have a key rotation policy to ensure that compromised keys will cease to be usable after a certain period. Replace and revoke a private key immediately if it is known to have been compromised to shut off access. | What is your key rotation policy? | **CNCF SSC:** SA: Have a key rotation policy |
| E.3.9 Ephemeral credentials | Reduce the number of potential entry points for a hacker, as well as the attack surface. | As an alternative to passwords, ephemeral credentials are randomly generated, short-lived access credentials that exist only for one session to authenticate and authorize privileged connections. These credentials are automatically issued as needed, so users do not have to input credentials when connecting and enable fine-grained permissions and automation of provisioning access tokens. | Do you use short-lived tokens for access management of machines and services such as CI/CD pipeline agents? Who uses these tokens and for what types of systems? | **CNCF SSC:** SA: Use short-lived/ephemeral credentials |
| E.3.10 Establish a root of trust | Provide the foundation for secure operations of a computing system | Roots of trust (RoT) are highly-reliable hardware, firmware, and software components that perform specific, critical security functions, such as storing the keys used for cryptographic functions and enabling a secure boot process. Many RoT are implemented in hardware so that malware cannot tamper with the functions they provide. The RoT is established from an offline source. | Do you have a root of trust can perform specific, critical security functions such as storing keys for cryptographic functions or enabling secure boot? How are these implemented? | **CNCF SSC:** SA: Follow established practices for establishing a root of trust |
| **DEPLOYMENT (D):** Tasks for identifying, analyzing, and addressing vulnerabilities in products in production/that have been deployed. | | | | |
| **D.1 Respond to/disclose vulnerabilities:** Tasks for identifying and addressing vulnerabilities in products and preventing similar ones from occuring in the future | | | | |
| D.1.1 Vulnerability analysis | Plan the resolution of discovered vulnerabilities | Enter discovered vulnerabilities into a defect management system. Analyze each vulnerability to gather sufficient information to plan its remediation. Plan a risk response and prioritization, such as by estimating the probability of exploitation and the impact if it is exploited. | How are discovered vulnerabilities recorded and triaged? How are the vulnerabilities prioritized for remediation? Is a severity rating, such as CVSS, used to aid prioritization? Is the access control to vulnerability information more restrictive than plain bug tracker defects? | **EO:** 4e(iv), 4e(viii)<br>**SSDF:** RV.2.1<br>**S2C2F:** SCA-5<br>**BSIMM:** CMVM2.2<br>**800-161:** SA-15<br>**Self-attestation:** 4 |
| D.1.2 Risk-based vulnerability remediation | Remediate vulnerabilities based upon a risk-based prioritization | Make risk-based decisions on whether to remediate a vulnerability or if the risk will be addressed another way (e.g. acceptance, temporary remediation, deferred remediation). Prioritize actions that will be taken. Deliver remediations via an automated and trusted delivery mechanism. | Can you tell me about your process for making decisions on which vulnerabilities should be remediated? What is your process for responding to reported incidents quickly? Can high-priority vulnerabilities prevent the shipment of the product? | **EO:** 4e(iv), 4e(v), 4e(viii)<br>**SSDF:** RV.2.2<br>**800-161:** SA-5, SA-11<br>**OSSF Scorecard:** Vulnerabilities, maintained<br>**Self-attestation:** 4 |

| Task Name | Objective | Definition | Question(s) | References |
| --- | --- | --- | --- | --- |
| D.1.3 Vulnerability disclosure | Aid organizations in responding to vulnerabilities to reduce the window of opportunity or attackers | Have a policy that invokes vulnerability disclosure and remediation, and organizational response. Implement the roles, responsibilities, and processes needed to support that policy, including a Product Security Incident Response Team (PSIRT) to handle responses to vulnerability reports and incidents. The disclosure program will require insight from internal stakeholders, such as legal, marketing, and public relations. | What is your process for addressing vulnerability disclosure and remediation with roles, responsibilities, and processes in place? How do you develop and release security advisories to your software acquirers? Do you publish a security policy (via a security.md file) to inform users about a vulnerability and how to report it? Do users know how to report vulnerabilities? Who handles PSIRT? | **EO:** 4e(viii)<br>**SSDF:** RV.1.3<br>**BSIMM:** CMVM1.1, CMVM2.1, CMVM3.7<br>**800-161:** SA-15<br>**S2C2F:** INV-2<br>**Self-attestation:** 4 |
| D.1.4 Vulnerability eradication | Proactively eradicate classes of vulnerabilities. | Review the software for similar vulnerabilities to eradicate a class of vulnerabilities not just the instance originally discovered. Proactively fix a class of vulnerabilities rather than waiting for the discovery of each vulnerability through automation, testing, or an external incident. Automation and custom rules in vulnerability discovery tools or compilers can be used. Patterns of remediation efforts should provide feedback into the secure SLDC. | Does your team have a process of proactively eradicating or reducing a whole class of vulnerabilities based upon previously-identified vulnerabilities or incidents? Are these patterns folded back into the SDLC? | **EO:** 4e(iv), 4e(viii)<br>**SSDF:** RV.3.3<br>**BSIMM:** CR3.3, CMVM3.1<br>**800-161:** SI-2<br>**Self-attestation:** 4 |
| D.1.5 Emergency artifact fix | Fix a zero-day vulnerability. | Zero-day vulnerabilities may not be fixed by an upstream maintainer in a desired timeframe. Implement an emergency response process to fix the vulnerability, re-build, deploy to your organization, and contribute the fix to the upstream maintainer. | How do you handle it when an external artifact has a zero-day vulnerability that is not being fixed by an upstream maintainer in a desired timeframe? | **S2C2F:** FIX-1 |
| D.1.6 Root cause analysis | Reduce the frequency of vulnerabilities in the future | Analyze vulnerabilities discovered throughout the process and in production to determine their root cause for being injected and not discovered earlier before production. Incident response feedback should be fed back to developers and be considered when evolve the SDLC. Record lessons learned. Identify patterns, such as security tasks in the SDLC needing to be followed. Discuss and disseminate these patterns to developers. Guide process correction if the current process is not being followed. | Do you analyze identified vulnerabilities to determine their root cause, including how the vulnerability was injected and, perhaps, how it escaped vulnerability detection efforts and made it into production? Do you communicate trends to developers? How can trends identified in root cause be factored into your SDLC process? | **EO:** 4e(ix)<br>**SSDF:** RV.3.1, RV3.2, RV.3.4<br>**BSIMM:** CP3.3, CMVM3.2 |
| **D.2 Monitor intrusions/violations:** Tasks for identifying malicious activity at runtime | | | | |
| D.2.1 System monitoring | Detect runtime product anomalies | Continuously monitor the running system to gather information for risk decisions, criticality analysis, vulnerability and threat analysis, incident response, policy non-compliance, and insider threat detection, including boundary protection, supply chain components, and supply chain information flow. | Are running systems continuously monitored to gather information for risk decisions, criticality analysis, vulnerability and threat analysis, incident response, and insider threat, including boundary protections of supply chain components and supply chain information flow? What is the process when an intrusion or violation is detected? Is there a QoS target? | **800-161:** CA-7 |
| D.2.2 Build process monitoring | Detect intruders in the build infrastructure | Do you monitor the build chain for unauthorized access and modifications? Have you completed an attack surface investigation of your build environment? What kinds of actions do you take to narrow the attack vectors? | Do you monitor the build chain for unauthorized access and modifications? Have you completed an attack surface investigation of your build environment? What kinds of actions do you take to narrow the attack vectors? | **CNCF SSC:** V: Validate runtime security of build workers; SA: Use short-lived workload certificates |